# Review of Sulphur Interaction based GaAs Surface Passivation and Its Potential Application in Magnetic Tunnel Junction based Molecular Spintronics Devices (MTJMSD)

*Pawan Tyagi*

Department of Mechanical Engineering,

NSF-CREST Center for Nanoscale Science and Technology
University of the District of Columbia, Washington DC 20008, USA.

Email of Corresponding Author: ptyagi@udc.edu

**Abstract** GaAs surface is characterized by a high density of surface states, which preclude the utilization of this semiconducting material for the realization of several advanced devices. Sulphur-based passivation has been found significantly useful in reducing the effect of dangling bonds. In this article first, the problem associated with GaAs surface has been discussed in a tutorial form. Secondly, the brief introduction of a wide variety of surface passivation methods was introduced. Sulphur passivation, the most effective surface state quenching method, has been elaborated. Thirdly, current trends in the field of surface passivation of GaAs surface has been discussed. Our discussion also focusses on utilizing GaAs and alloys for the molecular electronics and molecular spintronics and based on our insights in the GaAs (P. Tyagi, MRS Advances 2 (51), 2915-2920 2017) and molecular spintronics field (P. Tyagi, D. F. Li, S. M. Holmes and B. J. Hinds, J. Am. Chem. Soc. 129 (16), 4929-4938 (2007) and P. Tyagi, C. Riso, U. Amir, C. Rojas-Dotti and J. Martínez-Lillo, RSC Advances 10 (22), 13006, 2020) ).

**Introduction:** GaAs is one of the most important candidates electronic material for the fabrication of lasers, the base material for the quantum dots, and high-speed devices for the optical and mobile communication [1-3]. The full potential of GaAs, however, is limited by its unamenable surface electrical properties[2]. GaAs surface possesses myriad of surface state density at oxide/GaAs interface emanating from unsaturated dangling bonds of surface atoms and various surface defects[4, 5]. Which severely increases the surface recombination velocity (SRV), increase charge depletion region near the surface and pin the surface Fermi level near the mid-band gap [5].

Many attempts were made o deactivate the high surface state density using several elements and compounds for more than two decades but without any discovery of permanent solution of active surface problem[2, 6]. The sulfur (S) passivation has been the most successful surface treatment yielding a significant improvement in surface electrical properties and drew considerable attention [7-13]. Recently improved S passivation involved highly controlled atmosphere and high-temperature annealing yielded ideal surface electrical properties[14-16]. However, such a tardy surface passivation methodology involving a high temperature step apart from making process uneconomical will also affect the performance of other production steps. Due to this reason, efforts of the researchers are directed to develop a low temperature and less time-consuming surface passivation schemes [15, 17-19]. Recently advances have been made at this front by utilizing sulfur passivant in conjunction with other passivating elements or reagent for surface passivation [20, 21]. These new combined passivation schemes have produced higher degree improvement in surface electrical properties compared to conventional single-step sulfide treatment [21]. Moreover, the stability of the passivated surface also enhanced remarkably.

This article reviews GaAs surface passivation in the following manner. The problem associated with GaAs surface has been discussed. It is followed by the succinct introduction of important surface passivation schemes with the main emphasis on S passivation under separate subsections. Next section discusses the implication of surface passivation with other evolving fields such as molecular electronics and spintronics.

## 1. Problem with GaAs surface:

It is well known that the oxidized GaAs surface, the starting point for most of the devices, has high surface state density due to the missing cation and anion from the surface [22]. It has been observed that the ultra-high vacuum (UHV) cleaved (110) GaAs surface has no surface state induced levels within the band gap. However, UHV cleaved (100) surface exhibited nominal surface states density [23, 24]. This observation provided sufficient support for the concept that the very high surface states density was due to the interaction of foreign elements with GaAs surface. There exist several models which define the origin of surface state on the GaAs surface. According to one group of researchers, oxidized GaAs produced the surface state in the entire bandgap [18]. However, it has also been suggested that the high density of midgap surface states, which are mainly responsible for poor electrical properties of GaAs surface, are due to the presence of antisite defects [25, 26].

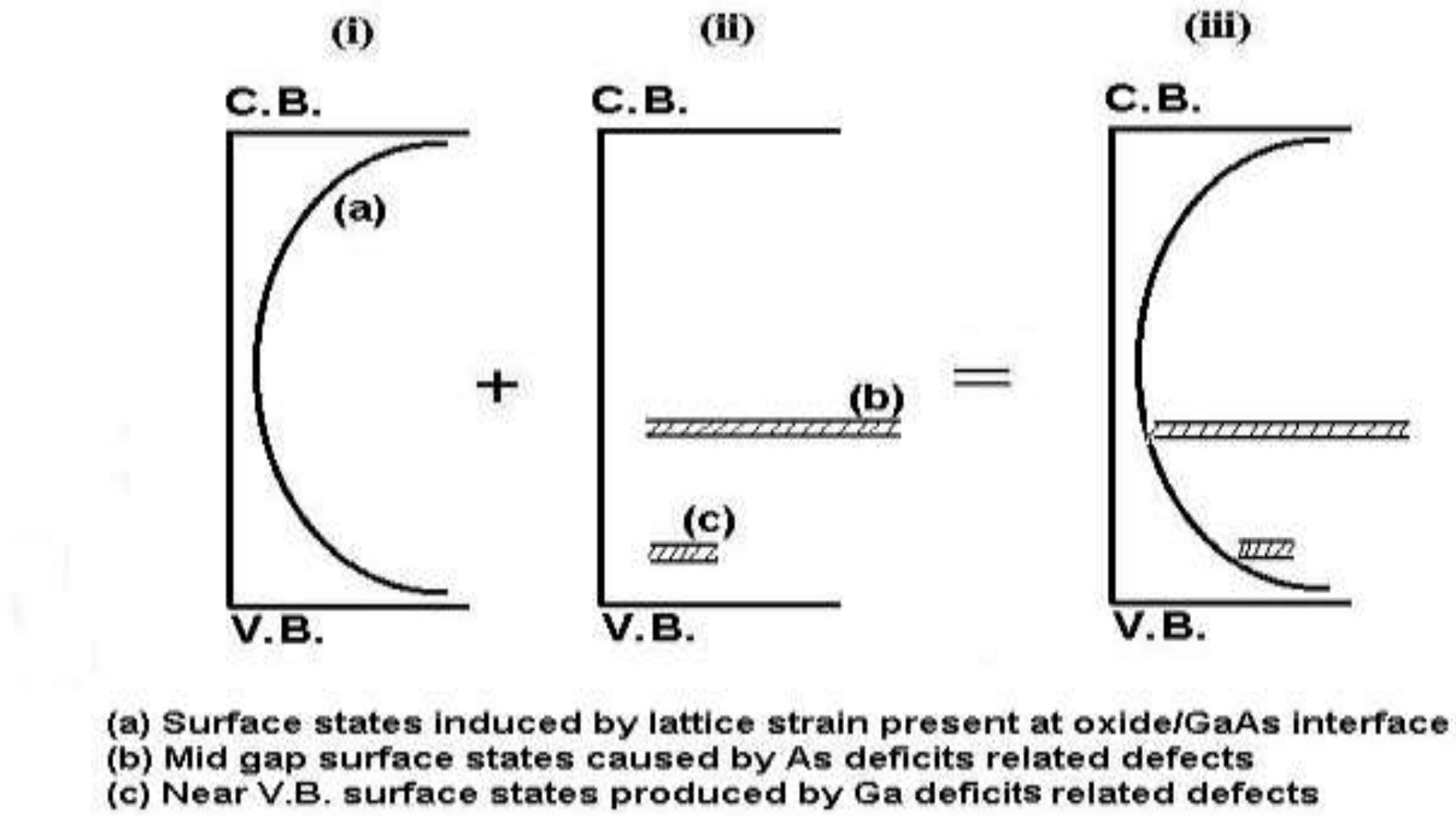

Fig.1 Appearance of surface states in band gap (i) surface states induced by lattice strain at oxide/GaAs interface, (ii) surface states induced by As and Ga deficits produced defects and (iii) cumulative presence of surface states density in energy band gap.

Spicer et al. [22] proposed that the release of adsorption energy of interacting oxygen (O) and various other metals atoms, introduced significant disorder at the GaAs surface layer. This produces local strain at oxide /GaAs interface and the presence of various types of defects like antisites and vacancies were reported [22]. According to them the high density of surface states presents around 0.8 eV below conduction band (CB), were due to As deficits possessing acceptor nature. Moreover, significantly high donor nature surface states near valance band (VB), but relatively quite small in comparison with midgap states, were believed to be caused by Ga deficit defects [22].

According to Besser et al. [26], surface states responsible for Fermi level pinning were introduced by arsenic (As) and gallium (Ga) antisite defects. They proposed that the singly and doubly ionized As antisites produced double donor level at 0.65eV and 0.9 eV below CB, respectively, and Ga antisite caused double acceptor levels near VB [25]. The pinning position of Fermi level was suggested to depend upon the equilibrium concentration of Ga and As antisites. They stated that the As-rich surface, energetically more favorable at room temperature and which appeared after GaAs etching in the acidic etchant, exhibited a very high density of As antisite defects. These As antisites defects were not compensated by the small amount of Ga antisites present, due to which the Fermi level was pinned by As antisite defects [26].

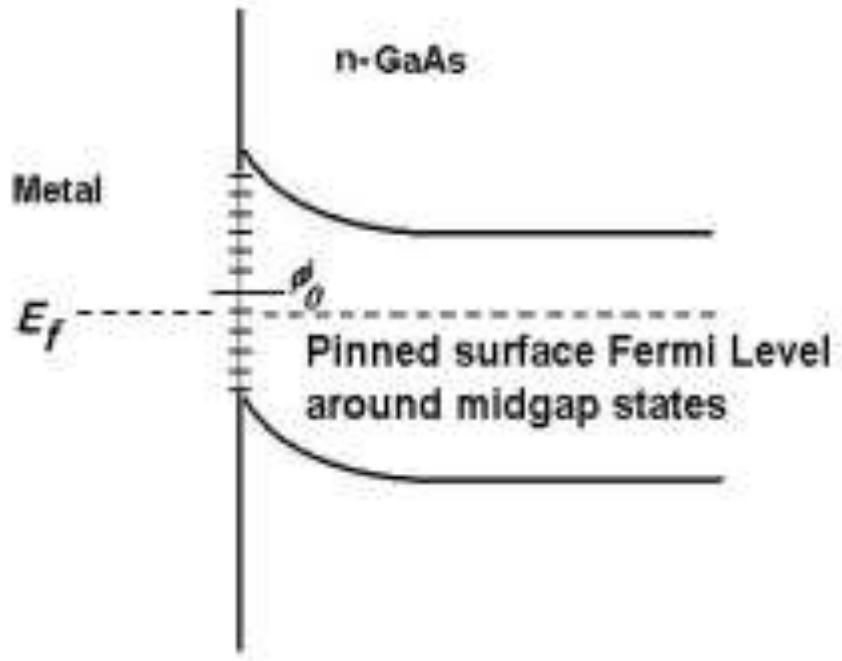

Fig.2. Model showing pinning of Fermi level around mid gap stated created by As deficit or As antisite defects.

The ensuing unified model that will be frequently used in this work is based on the review of the issue of surface states. There are mainly two sources of surface states in the GaAs band gap. The first source is local strain present at oxide/GaAs interface, which exhibited a parabolic distribution of surface states in the band gap. The second source of surface states are As and Ga deficit-induced defects like antisites, vacancies, etc. [22], which create high density of midgap surface states (due to As deficits or As antisites) and relativity low density of surface states near VB (due to Ga deficits or antisites) [25, 26]. Fig.1 shows the presence of surface states produced by local strain and these defects. The effect of above-described surface states on the surface band gap of GaAs is on a metal-semiconductor junction having insulator film in between is shown in Fig. 2. It clearly shows the Fermi level pinning approximately at midgap position. To eliminate these surface states from the energy gap passivation should be capable of removing oxides and deactivating antisite or As and Ga deficits defects. Moreover, the passivant should not create surface states due to its presence.

## 2. Surface Passivation of GaAs:

A large number of passivation schemes were employed to achieve unpinned Fermi level and surface state density at GaAs surface. The succinct list of major passivating elements and compounds comprises of $Ga_2O_3$ ($Gd_2O_3$) [27], nitrogen (N) [28], chlorine (Cl) [29], hydrogen (H) [30], polymerized thiophene [31, 32], phosphorous [33, 34], selenium (Se) [35], sodium diodecyldithiocarbamate (DODTC) [36], oxygen (O) [37, 38], fluorine (F) [21, 38-40] and most important of all sulfur (S) [7, 14, 15, 40, 41].

Kwo et al. [27] reported that the electron beam evaporation of $Gd_3Ga_5O_{12}$ garnet produced a mixed oxide film of $(Ga_2O_3)$1-x$(Gd_2O_3)$x having x% of $Gd_2O_3$. For x < 14% this film showed low leakage current, high break down strength, and good surface passivation characteristics. The explanation adduced to define the passivation capabilities of $(Ga_2O_3)_x$ $(Gd_2O_3)_{1-x}$ oxide layer was that Gd2O3 while exceeding 14% limit in oxide film either reduced the oxide vacancies or stabilize the 3+oxidation state of gallium oxide.

Nitrogen ion bombardment on GaAs (100) produced a strongly bonded GaN layer on the surface. GaN film, whose thickness was a function of bombarding ion energy, exhibited a reduction in the surface depletion region. Bombarded nitrogen also reduced the excess As and lattice disorder to yield improved surface properties [28].

Chlorine was used for GaAs surface passivation. Lu et al. [29] reported the realization of air-stable Cl terminated GaAs surface, exhibiting improved surface electrical properties. GaAs was dipped in 10% HCl for a few minutes to produce the Cl passivated surface. It was shown that GaCl bond was significantly more stable thermodynamically than As-Cl, Cl-H, and Ga-H bonds, due to which only GaCl bonds existed on GaAs surface after treatment.

Hydrogen passivation of GaAs surface, though extensively attempted in the late eighties and early nineties from both experimental and theoretical-view points, could not produce effective and stable deactivation of surface states. Its low-temperature stability precluded its application as a viable surface passivant [30].

Manorama et al. [31] have shown the passivation potentiality of a plasma-deposited polymer film on GaAs surface. They observed that polymer thiophene film was capable of reducing surface states and surface barrier height significantly. The effectiveness of thiophene as a passivant was validated by PL, capacitance-voltage (C-V), and Raman scattering measurements [32].

Beaudry et al. [33] reported the phosphorous (P) base passivation of GaAs surface. It was carried out by two methods, by actuating exchange reaction on GaAs surface in tertiary butyl phosphine vapor ambiance and, secondly by creating a direct thin GaP epitaxial layer. However, the growth temperature was kept significantly high. Passivation thus achieved was found stable over several months. These findings were in close agreement with the results presented elsewhere [34].

Selenium (Se) has also been used as a passivant for realizing flat band condition for surface band gap GaAs [35]. However, this process involved high temperature annealing which placed a significant amount of Se in GaAs surface after replacing bulk associated As. It was believed that the overall change in atomic arrangement on/in the GaAs surface was the reason behind unpinning of surface Fermi level [35].

A low temperature surface passivation scheme was suggested for the fabrication of a surface passivating insulator film through electro-deposition of sodium diodecyldithiocarbamate DODTC [36]. PL and Raman measurement validated the effectiveness of this novel method and confirmed the reduction in charge depletion region and unpinning of surface Fermi level to a significant extent. X-Ray Photoelectron Spectroscopy (XPS) study revealed that reactive sulphur present in DODTC, chemically bonded with surface As atoms to deactivate the surface state [36].

Surface passivation was also achieved by oxidizing GaAs surface through photo-oxidation method [37, 38]. Oxygen dissolved in flowing de-ionized water (DI) was charged on GaAs surface in the presence of white light. It was found that Fermi level was still pinned though photoluminescence intensity (PLI) increased significantly. Based on the experiment and theoretical defect model it was suggested that the PLI enhancement was possible without decreasing the surface state density [20,31]. Further research apropos of mechanism of PLI was thoroughly explored and above concept was discarded. A discussion on this subject will be made under the Mechanism section of this thesis.

Kim et al. [39] probed the passivation capabilities of gallium fluoride ($GaF_3$) on GaAs surface [39]. The $GaF_3$ film was directly deposited in a Molecular Beam Epitaxy (MBE) system. Electrical characterization of $GaF_3$/GaAs interface revealed the reduction in surface state density. Willston et al. [38] showed that fluoride ion ($F^+$) bombarded GaAs surface depleted surface As due to the formation of volatile arsenic fluorides. However, a gamut of fluorides comprises of GaF, $GaF_2$, and $GaF_3$ were present on the GaAs surface. It was observed that the stability of these three fluorides differed with respect to temperature. $GaF_3$ dominated at low temperature while GaF was the major fluoride at high temperature. More interestingly the GaAs surface was completely devoid of fluorides around 300°C. These studies are very helpful in understanding recent developments in the field of GaAs surface passivation utilizing F elements in conjunction with another passivant.

According to Jeng et al. [20], a drastic improvement in electrical properties of GaAs surface were witnessed after applying a joint passivation scheme involving S and F passivants. GaAs treated in phosphorous sulfide/ammonium sulfide [$P_2S_5/(NH_4)_2S$] and HF solution and followed by annealing at 300°C for 18 hrs showed a high degree improvement in electrical properties of GaAs surface while compared with individual treatment in $P_2S_5/(NH_4)_2S$ and HF solutions. It was expected that the formation of S fluoride such as $SF_6$ (having high binding energy of –291.8 kcal) was the most probable reason for this amelioration of electrical properties. However further research is required to create insight regarding the real mechanism of combined passivation. This finding suggests that the opportunities exist for improvement in surface properties of GaAs utilizing more than one passivant [35]. Tyagi [21] also attempted combined passivation approach by employing sulfide and fluoride ions from the solution.

Out of above-mentioned passivants sulfur(S) is the most successful element in ameliorating GaAs surface quality. S significantly decreases surface recombination velocity [25] and increases the PLI [42, 43]. Besides this surface barrier height become more sensitive towards metal work function [44, 45] after quenching of major part of the active surface states by S on GaAs surface [46]. The science and methodology of S passivation has improved significantly in past few years due to the extensive theoretical [47, 48] and experimental studies [7, 15, 49-52]. Complete realization of flat band energy gap, free from surface states, was reported using improved S passivation methods [45]. However, these improved S passivation techniques grossly depend upon processing at high temperature, ~400°C, under ultra-high vacuum (UHV) conditions [45, 52, 53].

Ensuing section is devoted to cover important aspects of S passivation methodologies, mechanisms, and other concerning factors from various viewpoints.

## 3. Sulphur passivation of GaAs:

Initially, S passivation was mainly carried out utilizing aqueous (aq.) solutions of sodium sulfide ($Na_2S$) [17, 41] and ammonium sulfide ($(NH_4)_2S$) (with or without dilution) [49, 50, 52, 54]. Further improvement in S passivation effectiveness was noticed by using a non-aqueous solvent to prepare a sulfide solution for passivation [7, 15, 53]. A change in the S charging method from sample dipping to electrochemical S charging also showed a definite improvement in electrical properties of GaAs surface [55-58]. In the quest for achieving stable S passivation and improved surface properties, many theoretical studies and experimental work were carried out to determine high-temperature characteristics of S passivated GaAs[8, 45]. Results of these studies showed that Ga-S bonds on GaAs provided very stable S passivation without providing any extra surface state in the energy band gap [48]. However, the As-S bonds were observed to be remarkably less stable and inferior from the viewpoint of their electrical properties. These facts prompted researchers to provide only Ga-S bonds on GaAs surface either by direct deposition [59-61] or by adding annealing treatment in addition to S passivation treatments [45].

### *3.1 Sulfur passivation methodologies*

(a) Passivation by aqueous sulfide solution: The solution utilized were $Na_2S$ (aq.) and $(NH_4)_2S$ (with or without dilution). The solution treatment to bring about surface passivation invariably start with surface cleaning (degreasing and deoxidation) of sample [17, 41]. The solution treatment to bring about surface passivation invariably start with surface cleaning (degreasing and deoxidation) of the sample [17, 41]. The cleaned sample have been dipped in solution at room temperature [62] or slightly higher temperature [62] for several minutes to many hours [5]. Sometimes, the cleaned GaAs is spin-coated with the sulfide solution [63]. Two main drying operations have been reported for sulfide solution treated samples, spin drying [46] and N2 drying [45]. The effects of processing parameters and solution characteristics of passivation properties are widely covered in ref. [10,64]. Here we will briefly discuss the effect of surface chemistry and temperature maintained during sulfidation. These two factors have a major impact on final bonding types on S passivated surface of GaAs.

Deoxidation of GaAs surface is an important treatment before sulfide passivation because it decides the surface chemistry before S passivation on GaAs sample. It is well known that Ga atoms of GaAs surface are highly dissolvable in acids such as HCl, $H_3PO_4$ , etc. and leave deoxidized As-rich surface after dipping GaAs sample in these acids [16]. However, ammonium hydroxide solution produces near stochiometric GaAs surface free from elemental As ($As^o$) [63]. Depending upon the relative amount of As and Ga quantity of As-S and Ga-S bonds changes on S treated sample [64, 65]. It was clearly shown that Na2S (aq.) solution produced the higher number of As-S bonds for a given As/Ga ratio while (NH4)2S has a greater tendency to yield more number of Ga-S bonds than Na2S (aq.) solution for a given As/Ga ratio [64, 66].

Characterization of sulfide solution-passivated GaAs surface revealed the following information. $Na_2S$ (aq.) and $(NH_4)_2S$ solution both increased the PLI [41]. However, it was reported that sodium sulfide treatment did not unpin the surface Fermi level [26] while $(NH_4)_2S$ do noticeable unpinning of surface Fermi level. Sodium sulfide solution passivation was observed to be very sensitive towards concentration of sulfide salt [67] while $(NH_4)_2S$ solution effectiveness was indifferent towards concentration factor [44]. S passivation effect vanished after thorough rinsing for both types of sulfide solution treatment [10]. $Na_2S$ (aq.) and $(NH_4)_2S$ solution were effective in reducing surface oxides, gallium oxide, and arsenic oxides [16, 41]. However, the dissolution rate of As° was significantly higher in $(NH_4)_2S$ solution as compared with $Na_2S$ (aq.) solution [50]. Sandroff et al. [40] noticed a clear distinction between nature of bonding S and GaAs surface for both the solutions. It was suggested that $Na_2S$ (aq.) solution provided $As_2S_3$ type compound on GaAs surface where S atom bridged between two As atoms while $(NH_4)_2S$ solution treated GaAs exhibited two S atoms staying between two As atoms [14, 15]. The mechanism of surface passivation will be thoroughly discussed later. The aq. $(NH_4)_2S$ solution passivation produced superior electrical properties compared to aq. $Na_2S$ passivation on GaAs surface.

(b) Passivation by non-aqueous sulfide solution : Sulfide solution treatment was more effective when sodium sulfide and ammonium sulfides were dissolved in a solvent with lower dielectric constant [68]. However, passivation methodology remains almost akin to aqueous sulfide solution passivation of GaAs surface. Properly degreased and deoxidized GaAs samples were immersed in non-aqueous sulfide solution at room temperature [7, 15, 68]. The residual solution was removed from the sample surface by spinning it at 1000 rpm all the time. The effectiveness of the application of non-aqueous solvents (alcohols of low dielectric constants such as ethyl alcohol ($C_2H_5OH$) [$\varepsilon$ = 25.3], isopropanol (i-$C_3H_7OH$) [$\varepsilon$ = 20.18], tertiary butanol (t-$C_4H_9OH$) [$\varepsilon$ = 12.47]) were justified by PL, XPS and Raman scattering studies [43, 68]. Results of PL and Raman studies are presented in Table 1.

**Table 1**: Results of PL depletion width (from Raman scattering study) and reactivity of S Changing with dielectric constant of solvent used. Data were taken from ref. [43].

| Passivation treatment in | Dielectric constant ($\varepsilon$) | PLI (a.u.) | Depletion width (nm) | Reactivity of S in solution (a.u) |
|---|---|---|---|---|
| No passivation | - | 1.0 | 33.6 | - |
| $(NH_4)S_x$ | - | 1.47 | 34.1 | 2.7 |
| $Na_2S+H_2O$ | 80.10 | 1.50 | 33.9 | 6.0 |
| $(NH_4)S_x$+ $C_3H_7OH$ | 20.18 | 1.70 | 26.2 | 80 |
| $(NH_4)S_x$+ t-$C_4H_9OH$ | 12.47 | 2.00 | 24.9 | 165 |
| $Na_2S$+ $C_3H_7OH$ | 20.18 | 2.30 | 24.2 | 8000 |
| $Na_2S$+ t-$C_4H_9OH$ | 12.47 | 3.58 | 23.6 | 212000 |

These results showed that the reactivity of S in solution was a strong function of the dielectric constant of the solvent. A decrease in the dielectric constant of solvent increased the reactivity of S in solution. The increase in PLI and a decrease in depletion width indicate a decrease in surface recombination velocity and unpinning of surface Fermi level from midgap states. The simultaneous improvement in both of the electrical properties indicated an overall change in surface states spectrum in energy band gap [43]. The performance of $(NH_4)_2S$ non-aqueous solution was inferior to the performance of $Na_2S$ non-aqueous solution (alcoholic solution), when the solvent of same dielectric constant was used, which has been explained by solution chemistry principal discussed elsewhere [15, 19]. Remarkable changes in the nature of surface bonds were also noticed during XPS study of non-aqueous $Na_2S$ solution treated GaAs surface (Table 2).

**Table 2:** Relative intensity of peaks corresponding to various type of bonds, surface coverage in monolayers (MLs) observed in core level XPS study of gallium (Ga3d) and arsenic (As3d) on GaAs surface with changing the value of solvent dielectric constant (ε). Data were taken from ref. [14].

| Treatment | ε | As-O (a.u.) | As-S (a.u.) | $As^o$ (a.u.) | Ga-O (a.u.) | Ga-S (a.u.) | S (a.u.) | Surface Coverage (MLs) |
|---|---|---|---|---|---|---|---|---|
| Untreated | - | 0.12 | 0 | 0.19 | 0.07 | 0 | 0 | 0 |
| $Na_2S+H_2O$ | 80.13 | 0.07 | 0.04 | 0.14 | 0.05 | 0 | 0.04 | 0.4 |
| $Na_2S+C_2H_5OH$ | 25.30 | 0.05 | 0.06 | 0.19 | 0.02 | 0.04 | 0.10 | 0.8 |
| $Na_2S+C_3H_7OH$ | 20.18 | 0.02 | 0.09 | 0.19 | 0.02 | 0.03 | 0.12 | 1.3 |

The data in Table 2 clearly shows that oxides removal and sulfur coverage was significantly better with a non-aqueous $Na_2S$ solution compared to aqueous solution. Excess As ($As^0$) removal was poor in case of non-aqueous $Na_2S$, but helped in justifying that elemental $As^0$ do not play any role in S passivation [63]. It was also observed that the formation of Ga-S bonds had also become possible by the utilization of alcoholic solvent [14]. The mechanism of S passivation will be presented later.

(c) Electrochemical S charging: Electrochemically charged S produced superior quality surface passivation over normal sulfide solution passivation [55-58]. In order to achieve surface passivation, the following methodology was adopted. After degreasing and deoxidation, as discussed for aqueous sulfide solution passivation methodology, the GaAs sample was so placed in a Teflon holder [55] or in a wax case [56] such that only one face of GaAs sample is exposed to S containing the electrolyte. Ohmic contact was established with the surface to be passivated prior to fixing the sample in the electrochemical cell [39]. Different types of S containing electrolytes were used: $Na_2S$-ethylene glycol [57], $(NH_4)_2S$ solution [57,57], a solution obtained from sequential mixing of propylene glycol with ammonia and hydrogen sulfide ($H_2S$) [56]. Supplying anodic current in milliamps [55] or microamps range [6] for few minutes charged sulfide ions. In the electrochemical cell, the counter electrode used was metal plate [55] and Pt [56]. The electrochemically charged GaAs surface was rinsed in ethanol and then blown in dry nitrogen gas by one research group [56].

XPS studies of electrochemically charged GaAs surface showed the presence of gallium sulfides ($Ga_2S_3$, GaS) and arsenic sulfide ($As_2S_5$) on it [56]. The passivation film, thus produced, was capable of providing a more stable deactivated surface in comparison with passivation film produced by ordinary sulfide solution treatment [55]. Moreover, electrochemical S passivation was significantly stable against water rinsing [57]. Presence of stable GaS and $As_2S_5$ ($As^{5+}$ state) were mainly responsible for the improvement in properties of electrochemically S passivated surface [57]. Additionally, it is well established that As in higher oxidation state provides better interface properties compared to $As^{3+}$ state [55].

(d) Deposition of GaS

The superior properties of Ga-S bonds on GaAs surface, prompted researches to deposit GaS directly on the suitably reconstructed surface. Realization of this process necessitated the availability of highly controlled systems and high substrate temperature to maintain required surface reconstruction and surface chemistry.

Direct deposition of GaS was successfully carried out by using gallium cluster ($[(t\text{-Bu})\ GaS]_4$) as a single source precursor in ultra-high vacuum conditions of the MBE system [60, 61, 69]. The substrate was cleaned by trisodium ethylaminoarsinic (TDMAAS) and Bisdimethyl aminochloroarsinic (BDMAASCl) [60]. During deposition of pure GaS, precursor cell was maintained at 120°C [60, 69][4,8]. The substrate temperature was kept in 350°C to 500°C temperature range. The substrate was at high temperature because it was essential to produce the required surface reconstruction, which played an important role in deciding the extent of improvement in electrical properties of GaS/GaAs interface. For instance, As rich (4x4) surface reconstruction produced $5x10^{10}$ $eV/cm^2$ surface states [60] while As rich (2x4) surface reconstruction yielded $1.8X10^{11}$ $eV/cm^2$ density of surface states [69] at GaS/GaAs interface. The deposited GaS film was reported to be amorphous and exhibited excellent surface morphology [60]. The GaS/GaAs interface exhibited excellent electrical properties only when the GaS film thickness was on the lower side since the excess amount of GaS produced high interface strain. The high lattice strain, present at interface facilitated the conversion of film nature from amorphous to polycrystalline, with a high density of dislocation density [60]. Moreover, the GaS film behaved like an insulator because it had a band gap of 3.5 eV [69]. Deposition of GaS produced an impressive enhancement in PLI and almost flattening of energy band gap near the surface. The passivation was also tested to be stable for at least two years [60].

(e) S passivation followed by annealing

Moriety et al. [43] used the following scheme to achieve almost complete unpinning of surface Fermi level in energy gap. First, the As-capped GaAs (100) samples were indium bonded to a tantalum or molybdenum holder before inserting the sample in UHV system. In the UHV system As decapping was carried out at 350 $^{o}$C. The As decapped samples were heated in 400-570°C range of temperature to produce c(2x8) As rich surface reconstruction and near stoichiometric (4x1) surface reconstruction. Thoroughly cleaned samples with c(2x8) and (4x1) starting surface reconstruction, were charged with molecular beam sulfur in an electrochemical cell. An anodic current of 0.5 mA was maintained for 5 minutes at room temperature. The sulfur charging step was followed by an annealing treatment from 450-500 $^{o}$C which yielded (2x1) surface reconstruction. XPS study of these samples confirmed the unpinning of surface Fermi level for the sample with (4x1) surface reconstruction [43].

It was shown that unpinning for (4x1) surface reconstruction was due to optimum atomic arrangement caused by S passivation and annealing. It was also shown that S atom had successfully replaced the bulk-associated surface As atoms, which was in agreement with the findings presented elsewhere [8, 70]. More importantly, the resultant atomic arrangement contained Ga-S bonds on the surface layer attached with partially or completely replaced bulk associated As [9, 45]. This experimental finding is in good accordance with the atomic arrangement, predicted for (2x1) reconstructed S passivated surface for engendering complete unpinning of surface Fermi level. In the case of c(2x8), As rich GaAs surface, it was found that replacement of bulk-attached As at surface with S was not carried out significantly after annealing of S passivated sample [45, 49]. This method again was energy intensive and required a high degree of control.

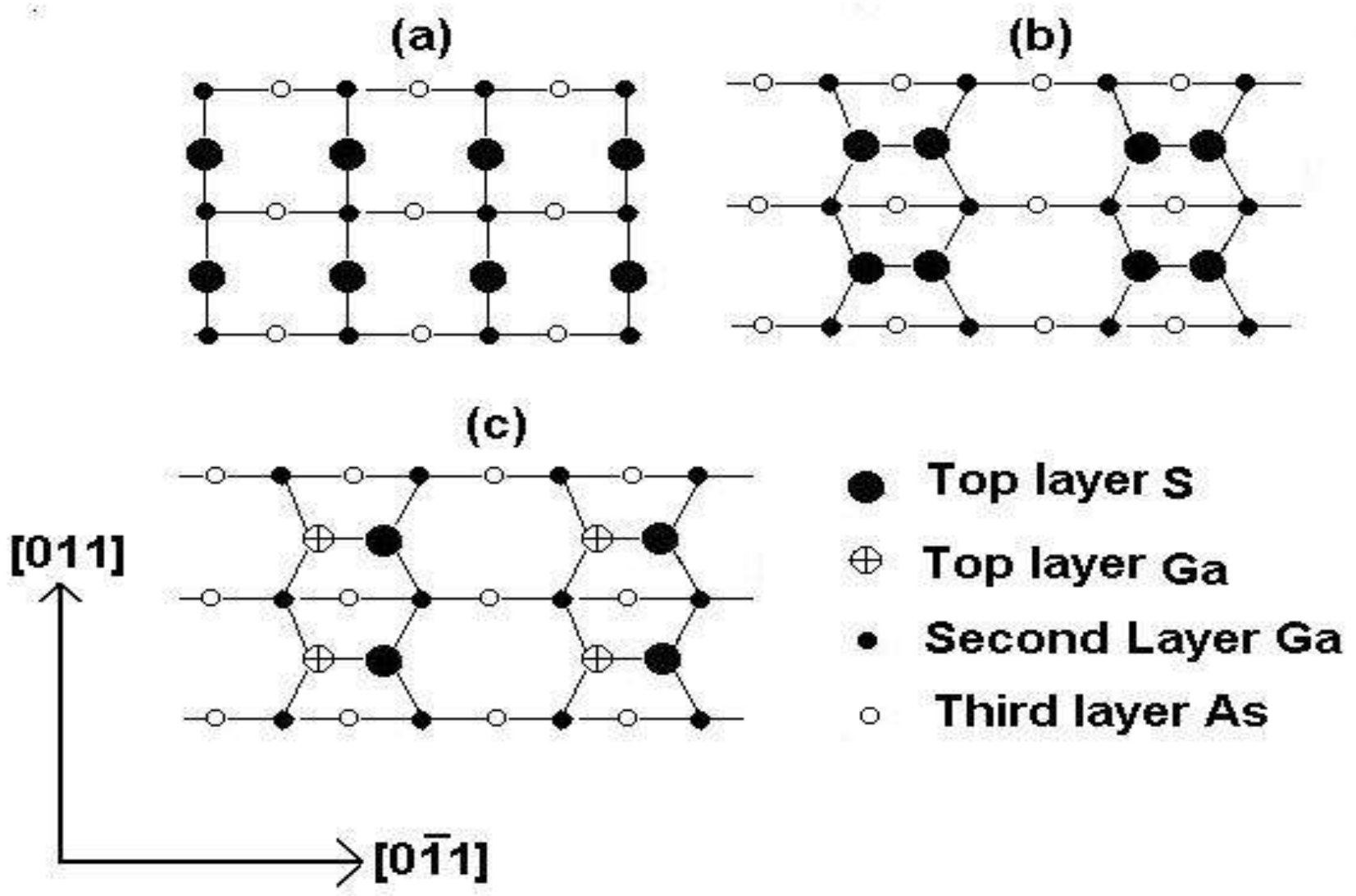

Fig. 3 (a) 1 ML of S on (1x1) surface reconstruction charged at RT, (b) 1 ML S charged and annealed produced (2x1) surface reconstruction, (c) ½ ML S charged and annealed with (2x1) surface reconstruction ref. [40].

## 3.2 *Atomic arrangement and surface reconstruction of passivated surface*

The atomic arrangement of S passivated surface mainly depended upon S coverage and availability of activation energy [46]. Three types of surface bonds, As-S, Ga-S, and S-S affects the surface state density in the energy band gap. The As-S bonds formation required lesser activation energy than that of Ga-S. The number of As-S and Ga-S, S-S bonds were found to be increasing with S coverage. More importantly, the existence of bonds S-S bonds is strongly dependent on the availability of S on GaAs than As-S, Ga-S [17].

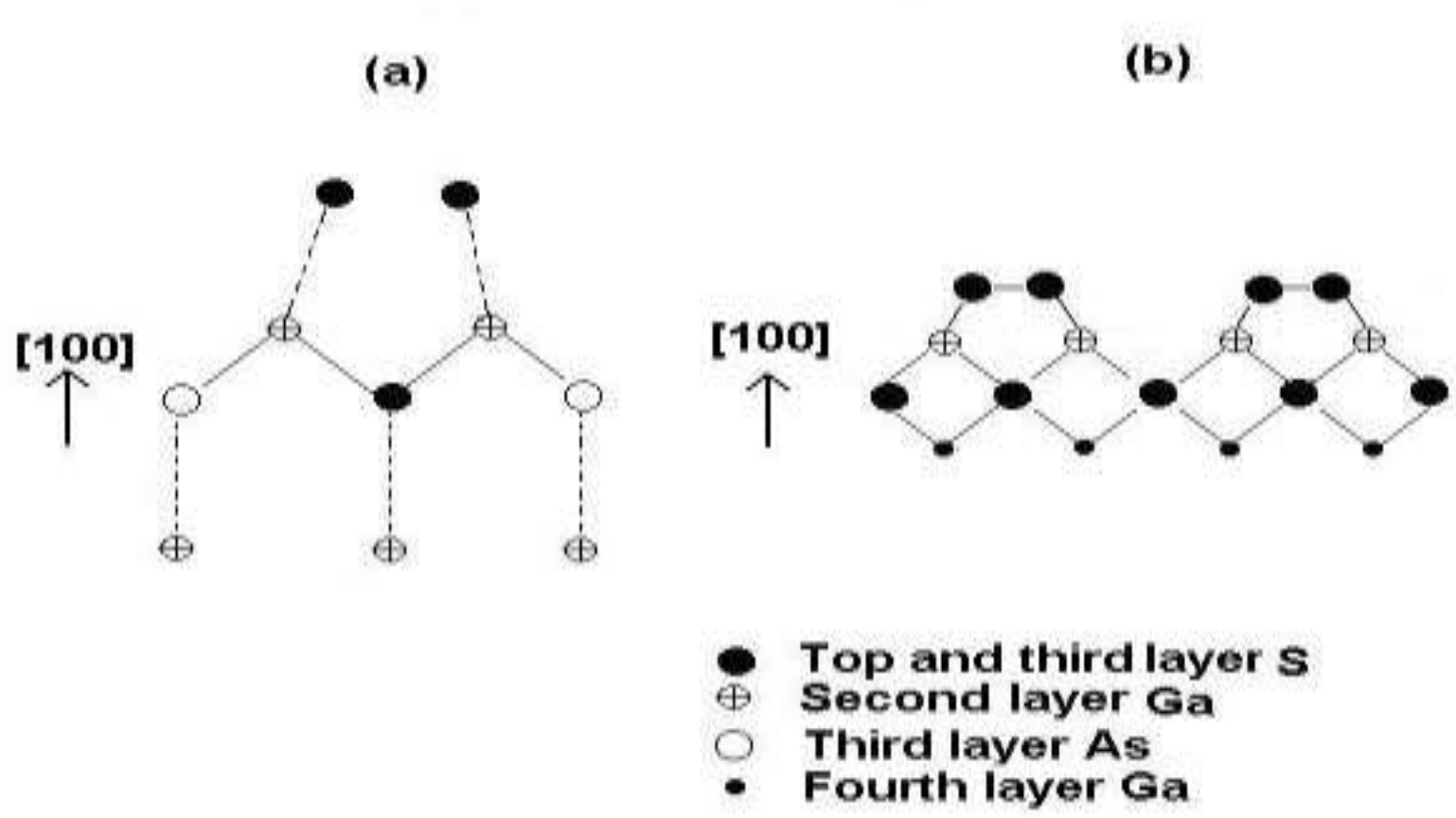

Fig. 4 (a) Optimal atomic arrangement obtained from theoretical study [44] (b) atomic arrangement proposed based on experimental study [70].

It was observed that S attached to As-S bonds releases itself upon increasing temperature and did not leave the GaAs surface until the temperature was high [56]. Freed S atoms in turn made bonds with surface Ga atoms and simultaneously start replacing bulk associated As atoms [8, 70]. According to the theoretical calculation, replacement of alternate bulk associated atoms by S on the surface with (2x1) reconstruction and Ga-S surface bonds is the optimal atomic arrangement for the realization of flat band condition which significantly matches with the atomic arrangement conceived based on experimental studies [45] Fig.4. The main difference is in the extent of removal of bulk associated As from the surface layer.

The impact of the presence of the above-described bond on the energy band diagram is following. First principle study suggested that As-S bond contained 2.25 electrons while for strong boding 2 electrons are needed, presence of excess 0.25 electrons, present in antibonding level weaken this bond. However, Ga-S bond contained exactly 2.0 electrons and had completely unfilled antibonding levels. Fig. 5 shows the electronic formation level due to As-S and Ga-S bond. Which lucidly showed that the appearance of new electronic level was due to the partially filled antibonding states of As-S bond. Whereas no level appeared in energy band gap due to Ga-S bond possessing filled boding state and completely vacant antibonding state. Electrical properties of S-passivated GaAs surface depend upon the final surface chemistry and atomic arrangement on the surface [6, 10, 11, 45, 71]. There exist significant differences in the nature of surface reconstruction and type of bonding after low-temperature sulfide solution passivation (T < 100 °C) and high-temperature sulfide passivation schemes. In near-room temperature sulfide passivation, surface chemistry mainly depends upon the type of methodology used [17, 63].

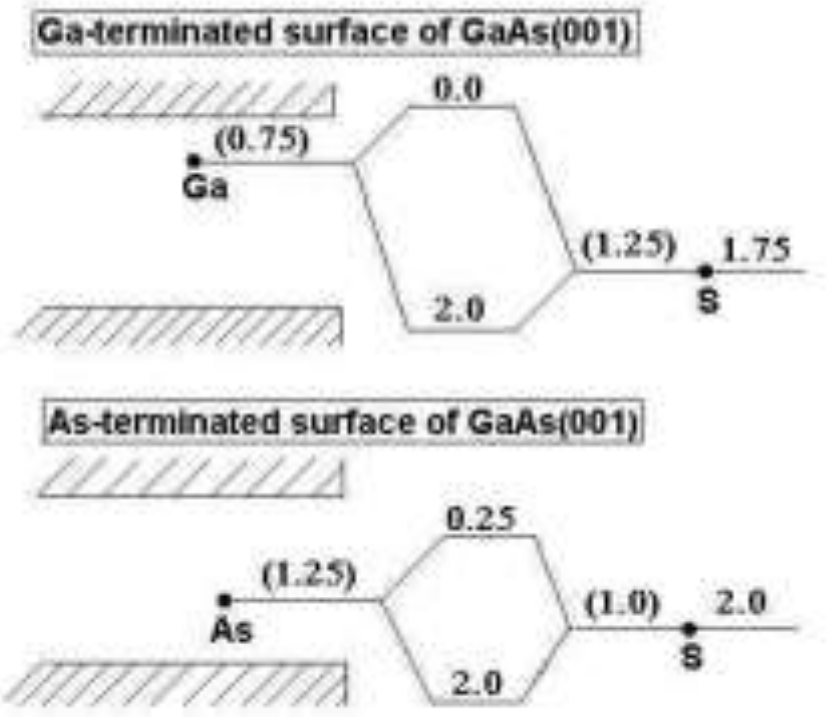

Fig. 5 Effect of As-S and Ga-S bonding on surface state after S passivation [43].

In the case of low temperature S passivation methods, sodium sulfide passivation mainly produced As-S type bonds on Ga terminated surface Ga-S, As-S and S-S bonds were observed after $(NH_4)_2S$ based sulfide treatment Fig. 6 [17]. The models presented in Fig. 3 for $(NH_4)_2S$ treated GaAs, have not incorporated the presence of Ga-S bonds, which were present on ammonium sulfide treated GaAs surface [17]. Presence of Ga-S bonds, along with As-S bonds, was validated by an XPS study conducted on non-aqueous $(NH_4)_2S$ solution treated GaAs surface [15]. It is quite apparent that there exists a need for a complete atomic model that must include As-S and Ga-S and Aso (refer to Table2).

Surface reconstruction, after room temperature S passivation of GaAs surface, was theoretically estimated to be (1x1). Though it matched with LEED results recorded on room temperature (RT) $H_2S$ passivated GaAs surface [70], it significantly differed from actual surface reconstruction (2x2) observed on $(NH_4)_2S$ treated surface [75,76]. Reason for this difference was the amount of S released by $H_2S$ and $(NH_4)_2S$ during passivation.[9, 43].

Electrochemical S passivation of GaAs surface evidently produced gallium sulfide ($Ga_2S_3$) and arsenic sulfide ($As_2S_5$) which were in higher oxidation state than gallium sulfide and $As_2S_3$ formed during simple sulfide solution treatment [55, 56]. Electrochemically passivated GaAs exhibited superior electrical properties and higher stability in comparison with normal sulfide passivation [55]. No appropriate atomic model was suggested for electrochemically passivated GaAs surface.

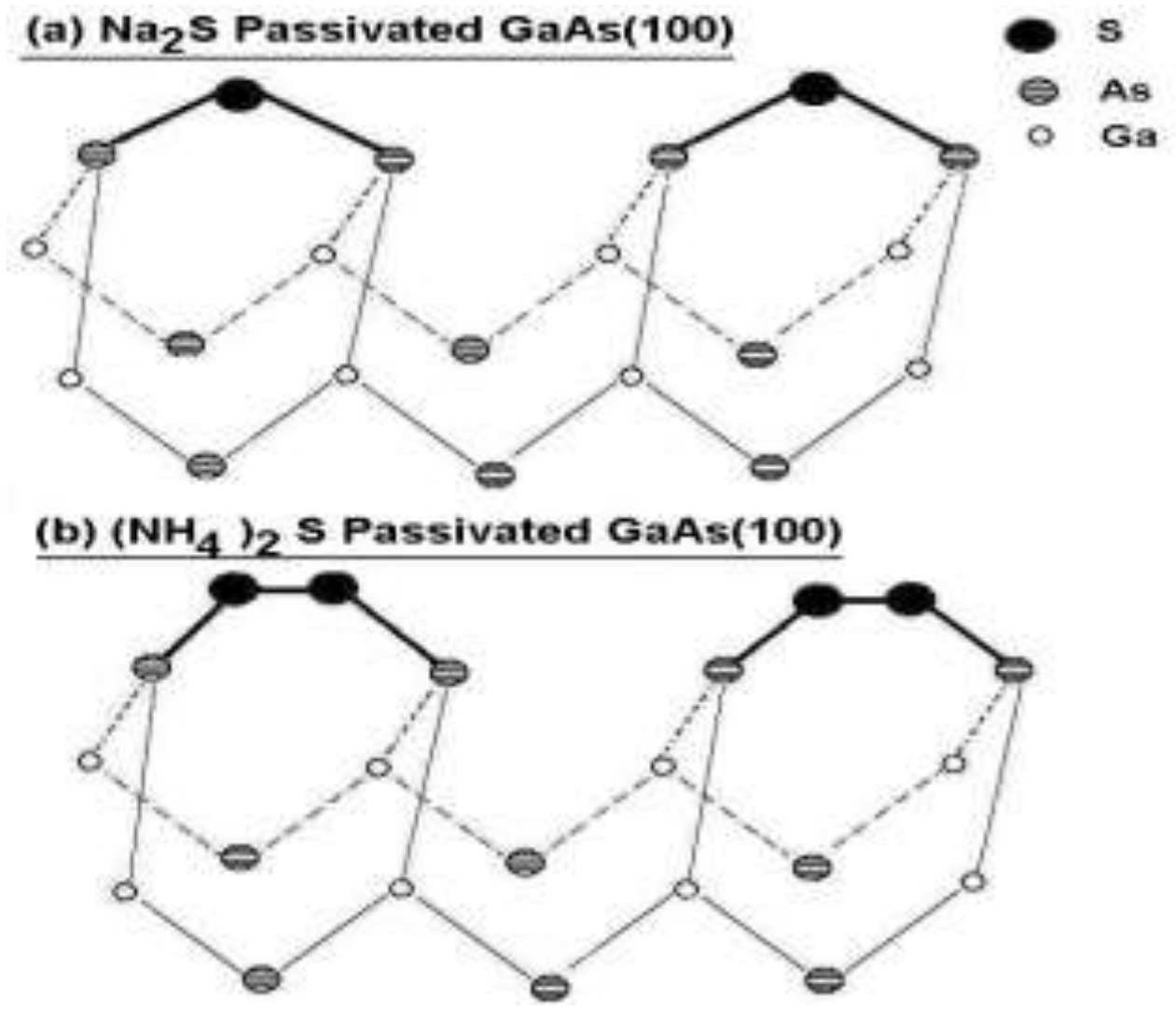

Fig. 7 Atomic arrangement of (a) $Na_2S$ and (b) $(NH_4)_2S$ passivated GaAs surface [37] Beeslov.

Close control over S passivation parameters becomes very necessary when efforts for realizing optimum atomic arrangements for complete unpinning of surface Fermi level. In order to grow thick GaS film on suitably reconstructed GaAs surface, controlled environment of the MBE system and high substrate temperature was employed [49]. By maintaining the substrate temperature in the 350-500°C range, As-rich surface reconstruction such as c(4x4) [60] and (2x4) [52] were established. On these surfaces, GaS was deposited to provide excellent electrical properties at the interface [59][3]. The choice of depositing GaS film, and not arsenic sulfide film, on GaAs surface, was based on Ga-S superior electrical properties and stability over As-S as discussed earlier [23]. GaS bonds are found to be stable up to ~530°C [13], while As-S bonds vanished before reaching 300°C [13, 72].

Another energy intensive passivation method in which S passivation was followed by annealing successfully produced surface energy gap with unpinned Fermi level [45]. This method, apart from producing GaS on the surface, also replaced a significant amount of substrate associated As atoms with S. This treatment yielded (2x1) surface reconstruction on S passivated surface [45]. Atomic model of S passivated, and annealed surface [70,76] was in close agreement with the result of theoretical study [44] as shown in Fig. 4. The significant difference in atomic models conceived based on the experimental and theoretical study was in the extent of substrate associated As replaced by S [59]. Suitable experimental studies are required to sort out this issue.

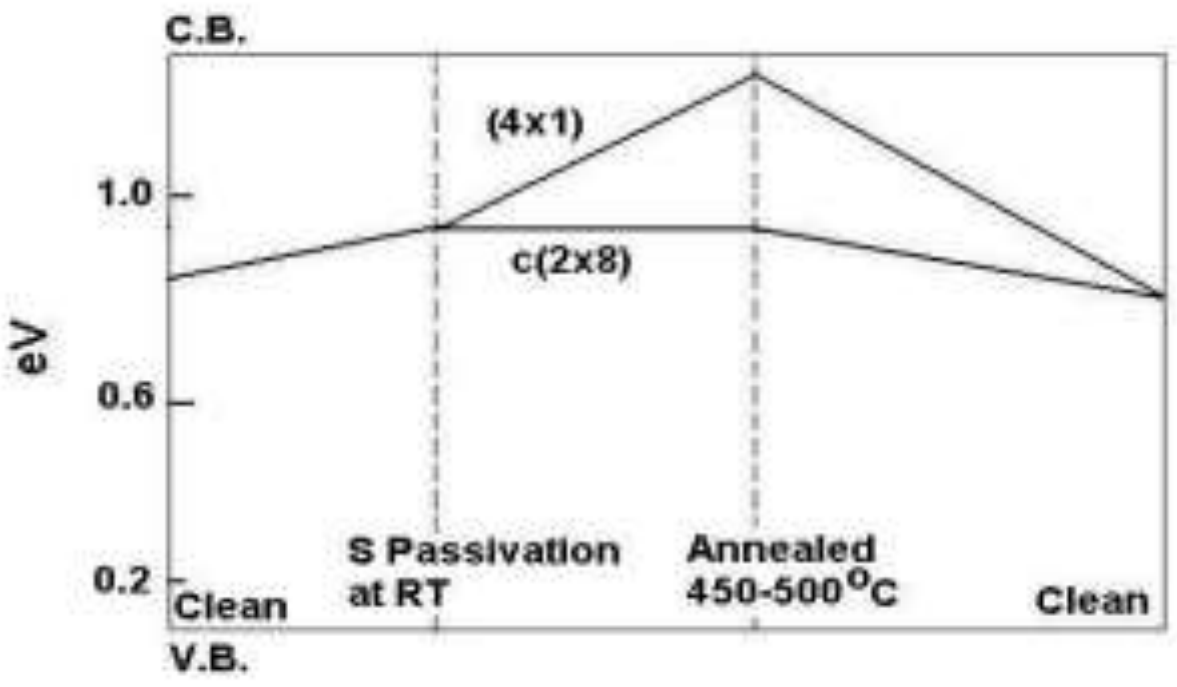

Fig. 8 Effect of surface reconstruction on the extent of unpinning of Fermi level. For As rich c(2x8) surface Fermi level is not unpinned even after annealing [39].

Moriety et al. [43] clearly showed the importance of starting surface chemistry and surface reconstruction. It was reported that As rich c(2x8) GaAs surface Fermi level remains unpinned due to inadequate Ga-S bonds formation and paucity of surface As atom replacement with S. Moreover, As dimer, which themselves create extra electronic states in energy gap, were also present after sulfidation and annealing treatment on surface c(2x8) reconstruction. Fig.8 shows that near stoichiometric GaAs surface with (4x1) surface reconstruction was observed with completely unpinned surface Fermi level after the same S passivation followed by annealing treatment [43].

Another surface reconstruction (2x6), which is also considered to be a stable one, was reported to have five S dimmers in a unit [73, 74]. On heating, this dimmer row breaks out the middle to yield (2x3) surface recombination [70]. Several researchers also showed that S passivated and annealed GaAs also tended to acquire S dimer attached with Ga atom [9, 75]. According to different studies, S dimer gave one electron to the Ga atom to provide a stable and neutral atomic arrangement on a passivated surface. Another critical aspect of S passivated GaAs surface is the direction of dimers present on it [76, 77]. S and As dimmers are observed to be oriented in [63] direction on S passivated and annealed around 450 °C. Heating beyond 550 °C, eliminated S and As dimers and gave rise to Ga dimers oriented in the [110] direction [76].

### 3.3 *Mechanism of S Passivation*

Several opinions exist on S passivation mechanism. Sandroff et al. [40, 41]claimed that the formation of protective $As_2S_3$ phase was the basis of S passivation. Another group of researchers suggested that amelioration of surface electrical properties were due to the band bending resulting from sulfide solution treatment [25, 41]. According to Hasegawa et al., sulfide treatment created a fixed negative charge near the surface, which improve the surface properties [41]. The most popular concept of S passivation is based on reduction in surface state density [15, 19].

The idea that $As_2S_3$ was the basis of S passivation was refuted by the findings presented in ref. [63]. In which two different S containing solutions gave equal enhancement to PLI even when $As_2S_3$ phase was absent in one case. Besser et al.[26] claimed that the band bending was induced by a sulfide solution on the GaAs surface, which pushed the electrons away from the surface or surface states affected the region. The increase in PLI was seen even when the Fermi level was still unpinned due to this reason [25]. A large number of references showed that the increase in PLI and Fermi level unpinning were observed after sulfide solution treatment forms the basis for the denial of the latter theory [15]. Hasegawa et al. [37] claimed that the increase in PLI was possible without decreasing surface state by the negative charge created by sulfide treatment on the GaAs surface. Experimental studies on S passivation shows that the source of surface states always decreased to a small or great extent after passivation treatment [50, 61]. On this basis, the concept of Hasegawa et al. [37] was found to be inconsistent with existing experimental results.

Aqueous $Na_2S$ solution improved PLI but Fermi level remain unpinned [26]. However, when non-aqueous $Na_2S$ solution was used for S passivation, unpinning of Fermi level was observed along with an increase in PLI [15]. It suggested that in the former case only a small part of the surface state was reduced, but in the latter case, the surface state reduction was significantly large. XPS results confirmed that the sources of surface states were greatly reduced by the non-aqueous sulfide treatment [15].

Low-temperature sulfidation is limited by the formation of As–S bond, which creates surface states within the band gap. Ga-S bond is more stable and does not produce extra surface states after passivation [78]. In order to harness the superior properties of Ga-S bond, the following two schemes were employed. First, direct deposition of GaS on suitably reconstructed As-rich surface in controlled UHV conditions was carried out [60]. While in another case, annealing treatment was added with S passivation of GaAs {Farrow, 1987 #277}[39].

Direct deposition of GaS from suitable precursor on a substrate, kept at high temperature in UHV, yielded extremely low surface state density at the interface. As there was no oxygen in the system, only termination of dangling bond on a reasonably good surface with already low surface states density, was left. Deposition of GaS deposition effectively passivated these remaining surface states to the extent that almost complete band bending was eliminated {Okamoto, 1999 #230}[4]. Sulfidation followed by annealing also yielded nearly flat band condition due to the following facts. Annealing treatment, apart from producing GaS bonds on the surface also replaced the surface associated As atoms with S {Conrad, 1996 #287}[59] to some extent to produce a very stable and electrically inert atomic arrangement. For this passivation treatment final surface chemistry determined by the XPS which was in close agreement with theoretically calculated surface chemistry for the complete unpinning of surface Fermi level from the mid-gap states. Reduction in surface states density depends upon methodology. S passivation relies on high-temperature treatment in a highly controlled environment for complete unpinning of the Fermi level.

## **4**. Present research trends and future scope:

The most successful sulfur passivation methods required a very controlled environment and annealing as an essential step in passivation scheme. Due to which it is challenging to assimilate the method into device fabrication routine. A low temperature and surface passivation method for the production of deactivated stable GaAs surface is the field of thirst. Recently few works have been reported regarding the use of more than one reagent to produce better surface electrical properties than single reagent sulfur passivation. Jeng et al. {Jeng, 1999 #231}[35] showed that the effectiveness of using both phosphorous sulfide/ ammonium sulfide (P2S5/(NH4)2Sx) solution and hydrogen fluoride (HF) solution on the barrier height enhancement of Ag/n-GaAs Schottky diode. A clear improvement in barrier height, more than single reagent sulfide passivation was exhibited. Researches have also shown that sulfide passivated GaAs manifested additional PLI enhancement after hydrogen annealing of already passivated GaAs surface . Furthermore, GaAs treated with sulfide solution followed by metal salt solution produced air stable surface passivation, which remained significantly stable for more than a year time. However, all the steps of this combined passivation were executed at room temperature in ambiance. The ability of combined surface passivation to produce superior surface electrical properties and more stable surface passivation has drawn considerable of researchers in this direction. This rather new stream of surface passivation is promising to yield more efficient, cost-effective less time consuming.

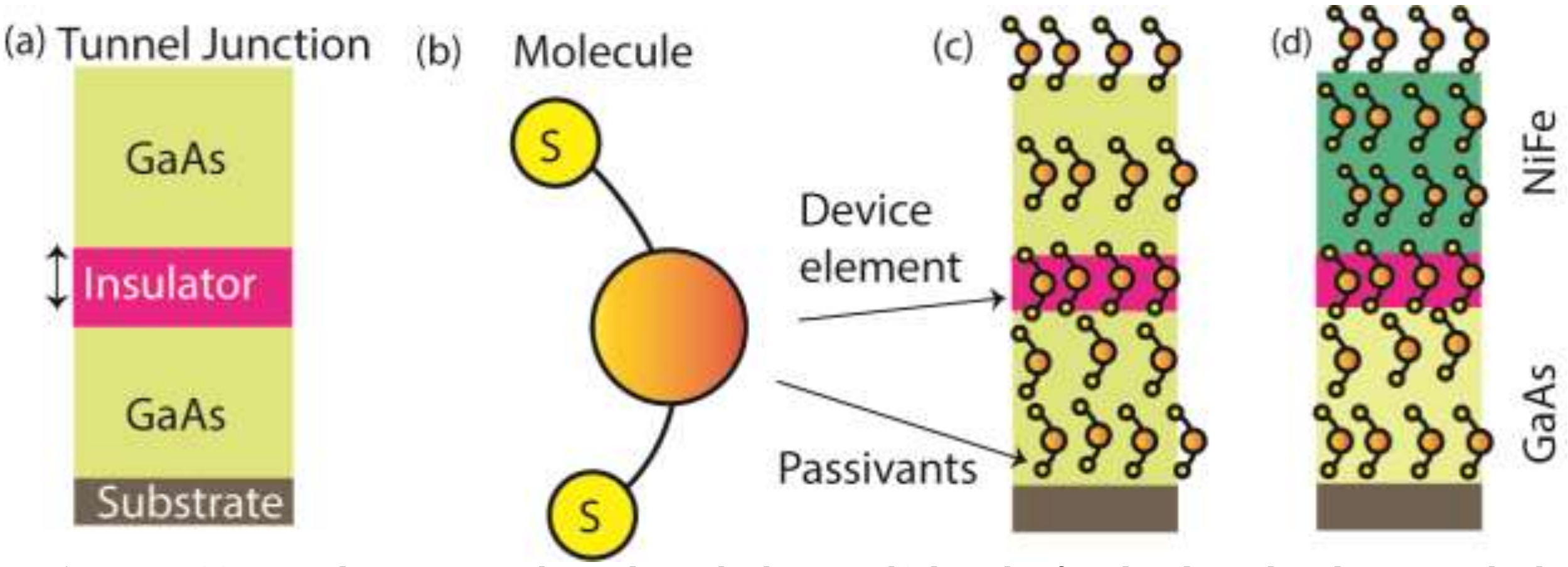

**Figure 9**: (a) Tunnel junction with insulator thickness< (b) length of molecule with at least two thiol (S) groups to interact with (c) GaAs as a device element and as a passivant (d) Tunnel junction with one GaAs electrode and one ferromagnetic electrode can be utilized. Ferromagnet can serve as a source of spin or spin detector.

**CONCEPT OF COMBINING SURFACE PASSIVATION WITH TUNNEL JUNCTION BASED MOLECULAR DEVICES**

The electrochemical sulfidation was effective in chemically bonding S with GaAs electrode. We propose that GaAs may be employed in a tunnel junction based molecular device scheme (Fig. 5) [18-26]. Under this approach a tunnel junction with the exposed edge can be produced in such a way that thickness of insulating gap (Fig.5a) is less the length of the molecule with at least two thiol ends (Fig. 5b)[25]. It is anticipated that the molecule bridging the insulating gap will serve as the device element and will govern the transport characteristics (Fig. 5c). Whereas, excess of molecules with thiol terminal will interact with exposed GaAs surface and will serve as the passivants (Fig. 5c). One of the GaAs can be replaced with a metallic electrode, e.g., NiFe ferromagnetic electrode (Fig. 5e) to produce molecular spintronics devices. Ferromagnetic electrodes in the molecular device can serve as the source of spin or detector of spin. Also, if GaAs is doped to acquire magnetic nature than the same molecule can also serve the additional role of establishing strong exchange coupling between the GaAs and ferromagnetic electrode leading to highly correlated materials, that can be termed as molecule-based metamaterials with unique physical properties[26-28]. In our prior study[26-28], paramagnetic molecules established unprecedented strong exchange coupling between two ferromagnetic electrodes. This exchange coupling resulted in intriguing current suppression and spin-dependent photovoltaic effect.

## Conclusion

In the present review, various features of surface-associated and problems and surface passivation methods and mechanism were discussed. Following are the main points emanate from the discussion made on active GaAs surface issue.

(i) An oxidized GaAs surface was characterized by two kinds of surface states. The first kind of surface states originated due to the lattice strain present at oxide semiconductor interface, and it possessed parabolic distribution in the band gap. The second type of surface states was produced by the As and Ga deficits which were produced by the release of adsorption energy of foreign atoms on the GaAs surface. The effectiveness of a surface passivation method is determined by its ability to passivate both kinds of surface states to the maximum extent.

(ii) A number of surface passivation schemes were executed to passivate the active GaAs surface; however, sulfide passivation was only observed to produced significant improvement in electrical properties and was widely researched.

(iii) Efficacy of sulfide passivation was found to very sensitive towards the type of methodology used to affect the surface state spectrum in the energy gap. Resultant surface chemistry of passivated GaAs surface depended upon initial surface chemistry, type of passivant, passivant charging method and post passivation treatment like annealing. For example, aqueous Na2S solution was only capable of reducing SRV without unpinning the surface Fermi level whereas alcoholic Na2S sulfide solution not only reduced the SRV but also freed the surface Fermi level from the mid-gap states. It appears that in former case passivant was only able to interact surface states mainly other than As deficits created mid-gap states whereas in later case sulfur atoms interacted with all the surface states and affected all the surface electrical properties of GaAs surface.

(iv) Recently combined surface passivation, using sulfur passivant with another reagent or element, was utilized to produce superior surface electrical properties at room temperature compared to simple sulfide solution treatment. Preliminary successes of combined surface passivation make this stream of surface passivation promising for yielding stable surface passivation and significant improvement in surface electrical properties of GaAs.

Acknowledgement

We gratefully acknowledge the funding support from National Science Foundation-CREST Award (Contract # HRD- 1914751), and Department of Energy/National Nuclear Security Agency (DE-FOA-0003945). PT also acknowledge support from UDC STEM center.

## REFERENCES:

[1] Sze, S. M., 1997, Modern Semiconductor Device Physics Wiley, John & Sons, Incorporated.

[2] Monier, G., Hoggan, P. E., Bideux, L., Paget, D., Mehdi, H., Kubsky, S., Dumas, P., and Robert-Goumet, C., 2019, "DFT and experimental FTIR investigations of early stages of (0 0 1) and (1 1 1) B GaAs surface nitridation," Applied Surface Science, 465, pp. 787-794.

[3] Williams, R. E., 1990, Modern GaAs Processing Methods, Artech House Publishers Boston.

[4] Raynal, P., Rebaud, M., Loup, V., Vallier, L., Roure, M., Martin, M., Barnes, J., and Besson, P., 2019, "GaAs WET and Siconi Cleaning Sequences for an Efficient Oxide Removal," ECS Journal of Solid State Science and Technology, 8(2), pp. P106-P111.

[5] Yngman, S., McKibbin, S. R., Knutsson, J. V., Troian, A., Yang, F., Magnusson, M. H., Samuelson, L., Timm, R., and Mikkelsen, A., 2019, "Surface smoothing and native oxide suppression on Zn doped aerotaxy GaAs nanowires," Journal of Applied Physics, 125(2), p. 025303.

[6] Cho, A. Y., 1976, "Bonding direction and surface-structure orientation on gaas (001)," Journal of Applied Physics, 47(7), pp. 2841-2843.

[7] Bessolov, V. N., Konenkova, E. V., and Lebedev, M. V., 1996, "Solvent effect on the properties of sulfur passivated GaAs," Journal of Vacuum Science & Technology B, 14(4), pp. 2761-2766.

[8] Conrad, S., Mullins, D. R., Xin, Q. S., and Zhu, X. Y., 1996, "Thermal and photochemical deposition of sulfur on GaAs(100)," Applied Surface Science, 107, pp. 145-152.

[9] Hou, X., Chen, X., Li, Z., Ding, X., and Wang, X., 1996, "Passivation of GaAs surface by sulfur glow discharge," Applied Physics Letters, 69(10), p. 1429.

[10] Paget, D., Bonnet, J. E., Berkovits, V. L., Chiaradia, P., and Avila, J., 1996, "Sulfide-passivated GaAs(001) .1. Chemistry analysis by photoemission and reflectance anisotropy spectroscopies," Physical Review B, 53(8), pp. 4604-4614.

[11] Paget, D., Gusev, A. O., and Berkovits, V. L., 1996, "Sulfide-passivated GaAs (001) .2. Electronic properties," Physical Review B, 53(8), pp. 4615-4622.

[12] Seo, J. M., Kim, Y. K., Lee, H. G., Chung, Y. S., and Kim, S., 1996, "Reduction of gap states of ternary III-V semiconductor surfaces by sulfur passivation: Comparative studies of AlGaAs and InGaP," Journal of Vacuum Science & Technology a-Vacuum Surfaces and Films, 14(3), pp. 941-945.

[13] So, B. K. L., Kwok, R. W. M., Jin, G., Cao, G. Y., Hui, G. K. C., Huang, L., Lau, W. M., and Wong, S. P., 1996, "Reordering at the gas-phase polysulfide passivated GaAs(110) surface," Journal of Vacuum Science & Technology a-Vacuum Surfaces and Films, 14(3), pp. 935-940.

[14] Bessolov, V. N., Konenkova, E. V., and Lebedev, M. V., 1997, "X-ray photoelectron spectroscopy study of GaAs(110) cleaved in alcoholic sulfide solutions," Journal of Vacuum Science & Technology B, 15(4), pp. 876-879.

[15] Bessolov, V. N., Konenkova, E. V., and Lebedev, M. V., 1997, "Sulfidization of GaAs in alcoholic solutions: A method having an impact on efficiency and stability of passivation," Materials Science and Engineering B-Solid State Materials for Advanced Technology, 44(1-3), pp. 376-379.
[16] Kang, M. G., Sa, S. H., Park, H. H., Suh, K. S., and Lee, J. L., 1997, "Sulfidation mechanism of pre-cleaned GaAs surface using (NH4)(2)S-x solution," Materials Science and Engineering B-Solid State Materials for Advanced Technology, 46(1-3), pp. 65-68.
[17] Bessolov, V. N., Lebedev, M. V., Binh, N. M., Friedrich, M., and Zahn, D. R. T., 1998, "Sulphide passivation of GaAs: the role of the sulphur chemical activity," Semiconductor Science and Technology, 13(6), pp. 611-614.
[18] Kang, M. G., and Park, H. H., 1998, "Effect of GaAs surface treatments using HCl or (NH4)(2)S-x solutions on the interfacial bonding states induced by deposition of Au," Thin Solid Films, 332(1-2), pp. 437-443.
[19] Bessolov, V. N., Lebedev, M. V., and Zahn, D. R. T., 1997, "Raman scattering study of surface barriers in GaAs passivated in alcoholic sulfide solutions," Journal of Applied Physics, 82(5), pp. 2640-2642.
[20] Jeng, M. J., Wang, H. T., Chang, L. B., Cheng, Y. C., and Chou, S. T., 1999, "Barrier height enhancement of Ag/n-GaAs and Ag/n-InP Schottky diodes prepared by P2S5/(NH4)(2)S-x and HF treatments," Journal of Applied Physics, 86(11), pp. 6261-6263.
[21] Tyagi, P., 2017, "GaAs(100) Surface Passivation with Sulfide and Fluoride Ions," MRS Advances, 2(51), pp. 2915-2920.
[22] Spicer, W. E., Chye, P. W., Skeath, P. R., Su, C. Y., and Lindau, I., 1979, "New and unified model for schottky-barrier and iii-v insulator interface states formation," Journal of Vacuum Science & Technology, 16(5), pp. 1422-1433.
[23] Lindau, I., Pianetta, P., Garner, C. M., Chye, P. W., Gregory, P. E., and Spicer, W. E., 1977, "PHOTOEMISSION STUDIES OF ELECTRONIC-STRUCTURE OF 3-5 SEMICONDUCTOR SURFACES," Surface Science, 63(1), pp. 45-55.
[24] Wilmsen, C. W., Kirchner, P. D., Baker, J. M., McInturff, D. T., Pettit, G. D., and Woodall, J. M., 1988, "CHARACTERIZATION OF PHOTOCHEMICALLY UNPINNED GAAS," Journal of Vacuum Science & Technology B, 6(4), pp. 1180-1183.
[25] Ohno, T., 1991, "Sulfur passivation of GaAs surfaces," Physical Review B (Condensed Matter)|Physical Review B (Condensed Matter), 44(12), pp. 6306-6311.
[26] Besser, R. S., and Helms, C. R., 1988, "EFFECT OF SODIUM SULFIDE TREATMENT ON BAND BENDING IN GAAS," Applied Physics Letters, 52(20), pp. 1707-1709.
[27] Kwo, J., Murphy, D. W., Hong, M., Opila, R. L., Mannaerts, J. P., Sergent, A. M., and Masaitis, R. L., 1999, "Passivation of GaAs using (Ga2O3)(1-x)(Gd2O3)(x), 0 <= x <= 1.0 films," Applied Physics Letters, 75(8), pp. 1116-1118.
[28] Delouise, L. A., 1993, "NITRIDATION OF GAAS(110) USING ENERGETIC N+ AND N2+ ION-BEAMS," Journal of Vacuum Science & Technology a-Vacuum Surfaces and Films, 11(3), pp. 609-614.
[29] Lu, Z. H., Chatenoud, F., Dion, M. M., Graham, M. J., Ruda, H. E., Koutzarov, I., Liu, Q., Mitchell, C. E. J., Hill, I. G., and McLean, A. B., 1995, "PASSIVATION OF GAAS(111)A SURFACE BY CL TERMINATION," Applied Physics Letters, 67(5), pp. 670-672.
[30] Herman, J. S., and Terry, F. L., 1993, "PLASMA PASSIVATION OF GALLIUM-ARSENIDE," Journal of Vacuum Science & Technology a-Vacuum Surfaces and Films, 11(4), pp. 1094-1098.
[31] Manorama, V., Bhoraskar, S. V., Rao, V. J., and Kshirsagar, S. T., 1989, "INTERFACIAL PROPERTIES OF N-GAAS AND POLYMER DEPOSITED BY PLASMA CHEMICAL VAPOR-DEPOSITION," Applied Physics Letters, 55(16), pp. 1641-1643.
[32] Rao, V. J., Manorama, V., and Bhoraskar, S. V., 1989, "PASSIVATION OF PINNED N-GAAS SURFACES BY A PLASMA-POLYMERIZED THIN-FILM," Applied Physics Letters, 54(18), pp. 1799-1801.
[33] Beaudry, R., Watkins, S. P., Xu, X. G., and Yeo, P., 2000, "Photoreflectance study of phosphorus passivation of GaAs (001)," Journal of Applied Physics, 87(11), pp. 7838-7844.
[34] Harrison, D. A., Ares, R., Watkins, S. P., Thewalt, M. L. W., Bolognesi, C. R., Beckett, D. J. S., and SpringThorpe, A. J., 1997, "Large photoluminescence enhancements from epitaxial GaAs passivated by postgrowth phosphidization," Applied Physics Letters, 70(24), pp. 3275-3277.

[35] Asai, K., Miyashita, T., Ishigure, K., and Fukatsu, S., 1995, "Electronic passivation of GaAs surfaces by electrodeposition of organic molecules containing reactive sulfur," Journal of Applied Physics, 77(4), pp. 1582-1586.
[36] Sawada, T., Hasegawa, H., and Ohno, H., 1987, "ELECTRONIC-PROPERTIES OF A PHOTOCHEMICAL OXIDE-GAAS INTERFACE," Japanese Journal of Applied Physics Part 2-Letters, 26(11), pp. L1871-L1873.
[37] Hasegawa, H., Saitoh, T., Konishi, S., Ishii, H., and Ohno, H., 1988, "CORRELATION BETWEEN PHOTOLUMINESCENCE AND SURFACE-STATE DENSITY ON GAAS-SURFACES SUBJECTED TO VARIOUS SURFACE TREATMENTS," Japanese Journal of Applied Physics Part 2-Letters, 27(11), pp. L2177-L2179.
[38] Williston, L. R., Bello, I., and Lau, W. M., 1992, "X-RAY PHOTOELECTRON SPECTROSCOPIC STUDY OF THE INTERACTIONS OF FLUORINE IONS WITH GALLIUM-ARSENIDE," Journal of Vacuum Science & Technology a-Vacuum Surfaces and Films, 10(4), pp. 1365-1370.
[39] Kim, K. H., Ishiwara, H., Asano, T., and Furukawa, S., 1988, "IMPROVEMENT OF THE INTERFACE PROPERTIES OF FLUORIDE GAAS(100) STRUCTURES BY POSTGROWTH ANNEALING," Japanese Journal of Applied Physics Part 2-Letters, 27(11), pp. L2180-L2182.
[40] Sandroff, C. J., Hegde, M. S., Farrow, L. A., Chang, C. C., and Harbison, J. P., 1989, "ELECTRONIC PASSIVATION OF GAAS-SURFACES THROUGH THE FORMATION OF ARSENIC SULFUR BONDS," Applied Physics Letters, 54(4), pp. 362-364.
[41] Sandroff, C. J., Hegde, M. S., and Chang, C. C., 1989, "STRUCTURE AND STABILITY OF PASSIVATING ARSENIC SULFIDE PHASES ON GAAS-SURFACES," Journal of Vacuum Science & Technology B, 7(4), pp. 841-844.
[42] McKay, L., and Wilson, L., "International Conference on Compound Semiconductor Manufacturing Technology," Proc. GaAs MANTECH Conference., GaAs MANTECH, St.Louis, MO, USA pp. 12-15.
[43] Moriarty, P., Murphy, B., Roberts, L., Cafolla, A. A., Hughes, G., Koenders, L., and Bailey, P., 1994, "PHOTOELECTRON CORE-LEVEL SPECTROSCOPY AND SCANNING-TUNNELING-MICROSCOPY STUDY OF THE SULFUR-TREATED GAAS(100) SURFACE," Physical Review B, 50(19), pp. 14237-14245.
[44] Carpenter, M. S., Melloch, M. R., Cowans, B. A., Dardas, Z., and Delgass, W. N., 1989, "INVESTIGATIONS OF AMMONIUM SULFIDE SURFACE TREATMENTS ON GAAS," Journal of Vacuum Science & Technology B, 7(4), pp. 845-850.
[45] Farrow, L. A., Sandroff, C. J., and Tamargo, M. C., 1987, "RAMAN-SCATTERING MEASUREMENTS OF DECREASED BARRIER HEIGHTS IN GAAS FOLLOWING SURFACE CHEMICAL PASSIVATION," Applied Physics Letters, 51(23), pp. 1931-1933.
[46] Xia, H., Lennard, W. N., Massoumi, G. R., Vaneck, J. J. J., Huang, L. J., Lau, W. M., and Landheer, D., 1995, "ABSOLUTE COVERAGE MEASUREMENTS ON SULFUR-PASSIVATED GAAS(100)," Surface Science, 324(2-3), pp. 159-168.
[47] Ren, S. F., and Chang, Y. C., 1990, "ELECTRONIC-PROPERTIES OF SULFUR-TREATED GAAS(001) SURFACES," Physical Review B, 41(11), pp. 7705-7712.
[48] Ohno, T., and Shiraishi, K., 1990, "First-principles study of sulfur passivation of GaAs(001) surfaces," Physical Review B (Condensed Matter)|Physical Review B (Condensed Matter), 42(17), pp. 11194-11197.
[49] Hirsch, G., Kruger, P., and Pollmann, J., 1998, "Surface passivation of GaAs(001) by sulfur: ab initio studies," Surface Science, 402(1-3), pp. 778-781.
[50] Hirayama, H., Matsumoto, Y., Oigawa, H., and Nannichi, Y., 1989, "REFLECTION HIGH-ENERGY ELECTRON-DIFFRACTION AND X-RAY PHOTOELECTRON SPECTROSCOPIC STUDY ON (NH4)2SX-TREATED GAAS (100) SURFACES," Applied Physics Letters, 54(25), pp. 2565-2567.
[51] Yablonovitch, E., Sandroff, C. J., Bhat, R., and Gmitter, T., 1987, "NEARLY IDEAL ELECTRONIC-PROPERTIES OF SULFIDE COATED GAAS-SURFACES," Applied Physics Letters, 51(6), pp. 439-441.
[52] Ha, J. S., Kim, S. B., Park, S. J., and Lee, E. H., 1995, "SURFACE-MORPHOLOGY OF (NH4)(2)S-X-TREATED GAAS(100) INVESTIGATED BY SCANNING-TUNNELING-MICROSCOPY," Japanese Journal of Applied Physics Part 1-Regular Papers Short Notes & Review Papers, 34(2B), pp. 1123-1126.
[53] Konenkova, E. V., 2002, "Modification of GaAs(100) and GaN(0001) surfaces by treatment in alcoholic sulfide solutions," Vacuum, 67(1), pp. 43-52.
[54] Yablonovitch, E., Skromme, B. J., Bhat, R., Harbison, J. P., and Gmitter, T. J., 1989, "BAND BENDING, FERMI LEVEL PINNING, AND SURFACE FIXED CHARGE ON CHEMICALLY PREPARED GAAS-SURFACES," Applied Physics Letters, 54(6), pp. 555-557.

[55] Hou, X. Y., Cai, W. Z., He, Z. Q., Hao, P. H., Li, Z. S., Ding, X. M., and Wang, X., 1992, "ELECTROCHEMICAL SULFUR PASSIVATION OF GAAS," Applied Physics Letters, 60(18), pp. 2252-2254.
[56] Yüzer, H., Dogan, H., Köroglu, J., and Kocakusak, S., 2000, "Analysis of sulfide layer on gallium arsenide using X-ray photoelectron spectroscopy," Spectrochimica Acta Part B: Atomic Spectroscopy, 55(7), pp. 991-996.
[57] Yota, J., and Burrows, V. A., 1993, "CHEMICAL AND ELECTROCHEMICAL TREATMENTS OF GAAS WITH NA2S AND (NH4)2S SOLUTIONS - A SURFACE CHEMICAL STUDY," Journal of Vacuum Science & Technology a-Vacuum Surfaces and Films, 11(4), pp. 1083-1088.
[58] Adlkofer, K., Tanaka, M., Hillebrandt, H., Wiegand, G., Sackmann, E., Bolom, T., Deutschmann, R., and Abstreiter, G., 2000, "Electrochemical passivation of gallium arsenide surface with organic self-assembled monolayers in aqueous electrolytes," Applied Physics Letters, 76(22), pp. 3313-3315.
[59] Macinnes, A. N., Power, M. B., Barron, A. R., Jenkins, P. P., and Hepp, A. F., 1993, "ENHANCEMENT OF PHOTOLUMINESCENCE INTENSITY OF GAAS WITH CUBIC GAS CHEMICAL VAPOR-DEPOSITED USING A STRUCTURALLY DESIGNED SINGLE-SOURCE PRECURSOR," Applied Physics Letters, 62(7), pp. 711-713.
[60] Okamoto, N., and Tanaka, H., 1999, "Characterization of molecular beam epitaxy grown GaS film for GaAs surface passivation," Materials Science in Semiconductor Processing, 2(1), pp. 13-18.
[61] Pelzel, R. I., Nosho, B. Z., Shoenfeld, W. V., Lundstrom, T., Petroff, P. M., and Weinberg, W. H., 1999, "Effect of initial surface reconstruction on the GaS/GaAs(001) interface," Applied Physics Letters, 75(21), pp. 3354-3356.
[62] Yuan, Z. L., Ding, X. M., Hu, H. T., Li, Z. S., Yang, J. S., Miao, X. Y., Chen, X. Y., Cao, X. A., Hou, X. Y., Lu, E. D., Xu, S. H., Xu, P. S., and Zhang, X. Y., 1997, "Investigation of neutralized (NH4)(2)S solution passivation of GaAs (100) surfaces," Applied Physics Letters, 71(21), pp. 3081-3083.
[63] Lunt, S. R., Ryba, G. N., Santangelo, P. G., and Lewis, N. S., 1991, "CHEMICAL STUDIES OF THE PASSIVATION OF GAAS SURFACE RECOMBINATION USING SULFIDES AND THIOLS," Journal of Applied Physics, 70(12), pp. 7449-7465.
[64] Shin, J., Geib, K. M., and Wilmsen, C. W., 1991, "SULFUR BONDING TO GAAS," Journal of Vacuum Science & Technology B, 9(4), pp. 2337-2341.
[65] Shin, J., Geib, K. M., Wilmsen, C. W., and Lillientalweber, Z., 1990, "THE CHEMISTRY OF SULFUR PASSIVATION OF GAAS-SURFACES," Journal of Vacuum Science & Technology a-Vacuum Surfaces and Films, 8(3), pp. 1894-1898.
[66] Geib, K. M., Shin, J., and Wilmsen, C. W., 1990, "FORMATION OF S-GAAS SURFACE BONDS," Journal of Vacuum Science & Technology B, 8(4), pp. 838-842.
[67] Fan, J. F., Oigawa, H., and Nannichi, Y., 1988, "THE EFFECT OF (NH4)2S TREATMENT ON THE INTERFACE CHARACTERISTICS OF GAAS MIS STRUCTURES," Japanese Journal of Applied Physics Part 2-Letters, 27(7), pp. L1331-L1333.
[68] Kang, M. G., and Park, H. H., 1999, "Effect of prepared GaAs surface on the sulfidation with (NH4)(2)S-x solution," Journal of Vacuum Science & Technology A, 17(1), pp. 88-92.
[69] Okamoto, N., Takahashi, T., Tanaka, H., and Takikawa, M., 1998, "Near-ohmic contact of n-GaAs with GaS/GaAs quasi-metal-insulator-semiconductor structure," Japanese Journal of Applied Physics Part 1-Regular Papers Short Notes & Review Papers, 37(6A), pp. 3248-3251.
[70] Berkovits, V. L., Bessolov, V. N., Lvova, T. N., Novikov, E. B., Safarov, V. I., Khasieva, R. V., and Tsarenkov, B. V., 1991, "FERMI-LEVEL MOVEMENT AT GAAS(001) SURFACES PASSIVATED WITH SODIUM SULFIDE SOLUTIONS," Journal of Applied Physics, 70(7), pp. 3707-3711.
[71] Chang, Y. C., Ren, S. F., and Aspnes, D. E., 1992, "OPTICAL ANISOTROPY SPECTRA OF GAAS(001) SURFACES," Journal of Vacuum Science & Technology a-Vacuum Surfaces and Films, 10(4), pp. 1856-1862.
[72] Tsukamoto, S., Ohno, T., and Koguchi, N., 1997, "Scanning tunneling spectroscopy and first-principles investigation on GaAs(001)(2x6)-S surface formed by molecular beam epitaxy," Journal of Crystal Growth, 175, pp. 1303-1308.
[73] Shimoda, M., Tsukamoto, S., Ohno, T., Koguchi, N., Sugiyama, M., Maeyama, S., and Watanabe, Y., 2000, "Stoichiometry study of S-terminated GaAs(001)-(2 x 6) surface with synchrotron radiation photoelectron spectroscopy," Japanese Journal of Applied Physics Part 1-Regular Papers Short Notes & Review Papers, 39(7A), pp. 3943-3946.

[74] Tsukamoto, S., and Koguchi, N., 1995, "SURFACE RECONSTRUCTION OF SULFUR-TERMINATED GAAS(001) OBSERVED DURING ANNEALING PROCESS BY SCANNING-TUNNELING-MICROSCOPY," Journal of Crystal Growth, 150(1-4), pp. 33-37.
[75] Shimoda, M., Tsukamoto, S., and Koguchi, N., 1998, "Photoelectron and Auger electron diffraction studies of a sulfur-terminated GaAs(001)-(2x6) surface," Surface Science, 395(1), pp. 75-81.
[76] Berkovits, V. L., and Paget, D., 1992, "OPTICAL STUDY OF SURFACE DIMERS ON SULFUR-PASSIVATED (001)GAAS," Applied Physics Letters, 61(15), pp. 1835-1837.
[77] Lu, Z. H., 1995, "Xanes studies of III-V semiconductor surface passivation," Progress in Surface Science, 50(1-4), pp. 335-345.
[78] Ke, Y., Milano, S., Wang, X. W., Tao, N., and Darici, Y., 1998, "Structural studies of sulfur-passivated GaAs (100) surfaces with LEED and AFM," Surface Science, 415(1-2), pp. 29-36.